\documentclass[twocolumn,english]{revtex4-1}
\usepackage[T1]{fontenc}
\usepackage[latin9]{inputenc}
\usepackage{color}
\usepackage{amsmath}
\usepackage{graphicx}
\usepackage{esint}

\makeatletter

\newcommand*\LyXThinSpace{\,\hspace{0pt}}
\providecommand{\tabularnewline}{\\}

\usepackage{babel}
\usepackage{amsfonts}

\usepackage{babel}

\makeatother

\usepackage{babel}
\begin{document}
\title{Quantum critical scaling of gapped phases in nodal-line semimetals}
\author{Geo Jose and Bruno Uchoa}
\affiliation{Department of Physics and Astronomy, University of Oklahoma, Norman,
Oklahoma 73019, USA}
\affiliation{Center for Quantum Research and Technology,University of Oklahoma,
Norman, Oklahoma 73019, USA}
\begin{abstract}
We study the effect of short range interactions in three dimensional
nodal-line semimetals with linear band crossings. We analyze the Yukawa
theories for gapped instabilities in the charge, spin and superconducting
channels using the Wilsonian renormalization group framework, employing
a large number of fermion flavors $N_{f}$ for analytical control.
We obtain stable non-trivial fixed points and provide a unified description
of the critical exponents for the ordering transitions in terms of
the number of order parameter components $N_{b}$ systematically to
order $1/N_{f}$. We show that in all cases, the dynamical exponent
$z=1$ in one loop, whereas $1/N_{f}$ corrections to various exponents
follow from the anomalous dimension of the bosonic fields only.
\end{abstract}
\maketitle

\section{Introduction}

Three dimensional (3D) nodal-line semimetals describe an interesting
class of fermionic systems where the valence and conduction bands
touch along manifolds of codimension $d_{c}=2$. Rather than forming
around points, such as in graphene \cite{Neto} or in Dirac and Weyl
semimetals \cite{Wan-1,Armitage}, the quasiparticles can appear around
closed lines in the 3D Brillouin zone, which are protected by symmetries
of the Hamiltonian \cite{Fang,Yang-sym,Chan-sym,Wang-sym,Chiu-1}.
Nodal-line semimetals have been originally predicted in a variety
of different contexts \cite{Arovas,Heikkila,Burkov,Xu,Mullen,Weng2,Kim,Weng}
and were more recently observed in different materials \cite{Xie,Bian,Bian2,Yan,Fu,Deng},
in photonic crystals \cite{Lu} and also in cold atom systems \cite{Song}.

In general, electron-electron interactions open up a myriad of new
possibilities for non-Fermi liquid behavior and various broken symmetry
phases in materials with nodal points or lines. Long range Coulomb
interactions in Dirac and Weyl semimetals have been shown to be isotropic,
marginal and lead to logarithmic corrections to physical quantities
at low energies \cite{Kotov,Sekine,Goswami,Hosur,Throckmorton,Gonsalez}.
In anisotropic systems, Coulomb interactions may lead to non-Fermi
liquid behavior over a wide range of energy scales \cite{Isobe,Moon,Herbut},
or else be irrelevant in the perturbative regime \cite{Huh,Wang-1,Yang-1}.
Short-range interactions, on the other hand, can lead to continuous
quantum phase transitions resulting in spontaneously ordered phases.
For instance, in the case of Dirac fermions in the honeycomb lattice,
the critical behavior was shown to belong to the Gross-Neveu-Yukawa
universality class \cite{Herbut-1,Assad}. 2D semi-Dirac fermions,
which exist at a topological phase transition between a semimetallic
and a gapped phase, where two Dirac cones merge \cite{Montambaux},
have unconventional quantum criticality \cite{Uryszek,Roy,Sur}. They
show correlation lengths that diverge along different directions with
distinct exponents, which may result in novel exotic superconductivity
with smectic order \cite{Uchoa}. In Weyl semimetals, short range
interactions lead either to a first-order phase transition into a
band insulator or else to a continuous transition into a symmetry
breaking phase \cite{Roy-1}.

\begin{figure}[b]
\begin{centering}
\includegraphics[width=0.84\linewidth]{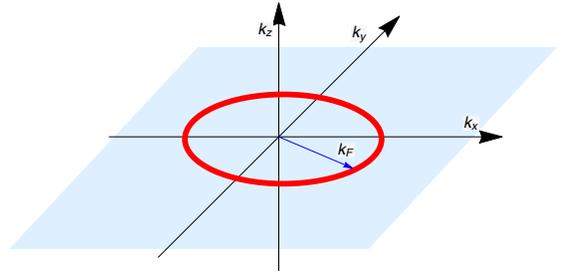}
\par\end{centering}
\caption{{\small{}{}Ring of nodal points in the $k_{z}=0$ plane resulting
from the spectrum of the free Hamiltonian (\ref{eq:hamiltonian}).
The light blue shaded region represents the plane of the nodal line.
Nodal loop of radius $k_{F}$ is shown in red.}}
\end{figure}

We contribute to this endeavor by investigating the effect of short-range
interactions on 3D nodal-line semimetals with linear band crossings.
We address the relatively general case where the nodal line forms
a closed ring or loop centered at the 3D Brillouin zone, as depicted
in Fig. 1. Due to the vanishing density of states at the nodal line,
the expected many-body instabilities occur through a quantum phase
transition separating the semimetallic regime from spontaneously broken
symmetry phases \cite{Nandikshore,Roy-2}. We investigate the universal
quantum critical scaling for instabilities in the spin, charge and
superconducting channels that produce fully gapped states.

Using a Wilson momentum shell renormalization group (RG) in the Yukawa
language, we analyze the interacting fixed points for the different
channels. We augment the action with a large number of fermionic flavors
$N_{f}$ for analytical control and derive their corresponding critical
exponents in leading $1/N_{f}$ order. In one loop, we show that the
mean field results are exact in the $s$-wave superconducting (SC)
channel, where vertex corrections vanish, whereas in both the charge
density wave (CDW) and spin density wave (SDW) orders we obtain finite
$1/N_{f}$ corrections. In all cases, the dynamical exponent 
\begin{equation}
z=1+O(N_{f}^{-2})\label{eq:z}
\end{equation}
in the regime where the radius of the nodal line is large compared
to all other energy scales\textcolor{blue}{.} The critical exponents
are summarized in a table, which is the main result of the paper.

As the outline of the paper, we introduce the Yukawa action of the
problem in section II. In section III, we discuss the Wilson RG scheme
for nodal-line semimetals, where we derive the RG equations for contact
interactions in the various channels. We then calculate the fixed
points and their respective critical exponents to leading order in
$1/N_{f}$. Finally, in section IV we present our conclusions.

\section{MODEL}

We consider the simplest non-interacting low energy Hamiltonian for
a nodal-line semimetal, which supports an isolated closed circular
loop of Fermi points with radius $k_{F}$. The nodal line is centered
around the origin of the Brillouin zone in the $k_{x}-k_{y}$ plane,
\begin{equation}
\mathcal{H}_{0}=\frac{k_{x}^{2}+k_{y}^{2}-k_{F}^{2}}{2m}\tau_{0}\otimes\sigma_{y}+v_{z}k_{z}\tau_{0}\otimes\sigma_{x},\label{eq:hamiltonian}
\end{equation}
where $\tau_{i}$ and $\sigma_{i}$ are Pauli matrices acting on the
spin and orbital/sublattice degrees of freedom. The four-component
spinor basis is defined as $\psi_{k}^{T}\equiv\left(\varphi_{1,\uparrow,\mathbf{k}},\varphi_{2,\uparrow,\mathbf{k}},\varphi_{1,\downarrow,\mathbf{k}},\varphi_{2,\mathbf{\downarrow,}\mathbf{k}}\right)$,
where $1,2$ denote pseudospin and $\uparrow,\downarrow$ spin quantum
numbers\textcolor{blue}{.} Lattice realizations of this model have
been proposed in different contexts, including hyperhoneycomb lattices
\cite{Mullen}, graphene networks \cite{Weng2} and cubic crystals
\cite{Roy-2}. Near the nodal loop, 
\[
\frac{k_{x}^{2}+k_{y}^{2}-k_{F}^{2}}{2m}\simeq v_{r}\tilde{k}_{r}
\]
where $\tilde{k}_{r}=\sqrt{k_{x}^{2}+k_{y}^{2}}-k_{F}$ and $v_{_{r}}=\frac{k_{F}}{m}$
is the radial Fermi velocity. Thus the quasiparticles disperse linearly
in all directions that are normal to the nodal line. The action for
the non interacting part is therefore given by
\begin{equation}
\mathcal{S}_{\psi}=\sum_{n=1}^{N_{f}}\int dk\ \bar{\psi}_{n,k}\left[-ik_{0}+\mathcal{H}_{0}\right]\psi_{n,k}
\end{equation}
where $k=\left(k_{0},k_{x},k_{y},k_{z}\right)$ is the four-momentum
vector in 3+1 dimensions, with $k_{0}$ the frequency and $\int dk\equiv(2\pi)^{-4}\int dk_{0}dk_{x}dk_{y}dk_{z}.$
$\psi_{n,k}$ represents the fermion field carrying a flavor index
$n$, with $N_{f}$ the number of fermionic flavors, which are treated
as a degeneracy.

To study the quantum critical behavior of the nodal-loop system, we
use a Hubbard-Stratanovich decomposition of the four-fermion interaction
into appropriate channels and study the resulting Gross-Neveau-Yukawa
theories. Short range interactions can lead to mass terms of the form
$\sum_{i=1}^{N_{b}}M_{i}\Gamma_{i}$, where $\Gamma_{i}$ are all
possible $4\times4$ matrices that anticommute with the noninteracting
Hamiltonian (2). In this class of phase transitions, the mass term
describes the spontaneous chiral symmetry breaking of the system across
a quantum critical point into a gapped phase. The Yukawa coupling
term in the action can then be generically written as
\begin{equation}
\mathcal{S}_{\phi\psi}=g\sum_{n=1}^{N_{f}}\sum_{j}^{N_{b}}\int dkdq\,\phi_{q}^{j}\left(\bar{\psi}_{n,k}\Gamma_{j}\psi_{n,k-q}\right)\label{eq:Yukawa term}
\end{equation}
where 
\begin{equation}
\Gamma_{j}=\tau_{j}\otimes\sigma_{3},\label{eq:Gammaj}
\end{equation}
with $j=0,1,2,3$ are the only four possible mass terms that lead
to gapped phases in the considered Hilbert space.

CDW and SDW instabilities correspond to staggered patterns of charge
and spin in the pseudospin space. The effective single particle interaction
Hamiltonian that describes those instabilities is of the form 
\begin{equation}
\mathcal{H}_{CDW}=\phi^{0}\tau_{0}\otimes\sigma_{3},\label{CDW}
\end{equation}
 and 
\begin{equation}
\mathcal{H}_{SDW}=\boldsymbol{\phi}\cdot\boldsymbol{\tau}\otimes\sigma_{3}\label{SDW}
\end{equation}
respectively, where $\boldsymbol{\phi}=\left(\phi^{1},\phi^{2},\phi^{3}\right)$
is a vector order parameter and $\boldsymbol{\tau}=\left(\tau_{1},\tau_{2},\tau_{3}\right)$.
Clearly, these terms anticommute with the noninteracting Hamiltonian
(2), leading to gaps in the spectrum. To identify the vertex of the
$s$-wave superconducting case, it is convenient to double the size
of the Hilbert space and introduce the 8-component Nambu spinor basis
$\Psi_{k}^{T}=\left(\varphi_{1,\uparrow,\mathbf{k}},\varphi_{2,\uparrow,\mathbf{k}},\varphi_{1,\downarrow,\mathbf{k}},\varphi_{2,\mathbf{\downarrow,}\mathbf{k}},\varphi_{1,\downarrow,-\mathbf{k}}^{\dagger},\varphi_{2,\downarrow,-\mathbf{k}}^{\dagger},\varphi_{1,\uparrow,-\mathbf{k}}^{\dagger},\varphi_{2,\mathbf{\uparrow,}-\mathbf{k}}^{\dagger}\right)$.
In this basis, the non-interacting part of the Hamiltonian can be
written as 
\begin{equation}
\mathcal{H}_{0}=\frac{k_{x}^{2}+k_{y}^{2}-k_{F}^{2}}{2m}\eta_{0}\otimes\tau_{0}\otimes\sigma_{y}+v_{z}k_{z}\eta_{0}\otimes\tau_{0}\otimes\sigma_{x},\label{Ho2}
\end{equation}
where the Pauli matrices $\eta_{i}$ ($i=0,1,2,3$) act in the Nambu
space. The fully gapped $s$-wave pairing term that anticommutes with
(\ref{Ho2}) has the form $\mathcal{H}_{SC}=\boldsymbol{\phi}\cdot\boldsymbol{\eta}_{\perp}\otimes\tau_{0}\otimes\sigma_{3}$
where $\boldsymbol{\phi}=(\phi_{1},\phi_{2})$ is a vector whose components
are defined in terms of the real and imaginary parts of the order
parameter, and $\boldsymbol{\eta}_{\perp}=(\eta_{1},\eta_{2}).$ One
can see that the spin space is redundant in this basis, as it corresponds
to two identical $4\times4$ copies of the Hamiltonian. Therefore,
we drop for convenience the $\tau_{0}$ matrices in $\mathcal{H}_{0}$
and $\mathcal{H}_{SC}$ and absorb the spin as a degeneracy. In that
case, the pairing term has the form 
\begin{equation}
\mathcal{H}_{SC}=\boldsymbol{\phi}\cdot\boldsymbol{\eta}_{\perp}\otimes\sigma_{3}.\label{HSC}
\end{equation}
The Yukawa vertex that follows from this term has the same form of
Eq. (\ref{eq:Gammaj}) for $j=1,2$ if one performs the substitution
$\eta_{i}\leftrightarrow\tau_{i}$. As expected, the pairing term
is dual to the antiferromagnetic XY model.

In all cases, we can identify the different mass terms that correspond
to distinct channels of instabilities in the charge, spin and $s$-wave
superconducting states with a Yukawa vertex of the form (\ref{eq:Gammaj})
written in some appropriate basis. The different ordered states encoded
in the generic Yukawa coupling (\ref{eq:Gammaj}) are
\begin{equation}
j=\begin{cases}
0 & \text{CDW}\\
1,2 & \text{SC}\\
1,2,3 & \text{SDW}
\end{cases}.
\end{equation}
Here, $N_{b}=1,2$,$3$ describes respectively the number of bosonic
field components in the CDW, SC and SDW cases, respectively.

Other possible emergent mass terms leading to fully gapped states
are allowed if one enlarges the size of the Hilbert space. For instance,
in 2D Dirac fermions on the honeycomb lattice, an anomalous quantum
Hall (AQH) state is in principle allowed \cite{Raghu,Christou} when
one reverses the sign of the mass term across opposite valleys. In
nodal-line semimetals, where the concept of valleys is ill defined,
the reversal of the sign of the mass in the AQH state happens continuously
along the nodal line. Due to particle-hole symmetry, which keeps the
mass purely imaginary and hence topological, this state does not produce
a fully gapped state, but rather a Weyl semimetal, with Fermi arcs
connecting a discrete number of gapless points of the nodal line,
where the mass term changes sign \cite{Kim-Hall}. Although this is
an interesting state, it is highly dependent on the microscopic details
of the lattice model, such as the number of nodes of the mass along
the nodal line \cite{Kim-Hall,Okugawa}, and will be considered elsewhere.
Here, we will restrict our analysis to isotropic instabilities in
the considered Hilbert space that fully gap the nodal line.

The free part of the bosonic action can be written as
\begin{equation}
\mathcal{S}_{\phi}=\sum_{j=1}^{N_{b}}\int dq\phi_{-q}^{j}\left(c_{0}^{2}q_{0}^{2}+c_{1}^{2}(q_{x}^{2}+q_{y}^{2})+c_{2}^{2}q_{z}^{2}+m_{\phi}^{2}\right)\phi_{q}^{j},\label{Sphi}
\end{equation}
in which we have also included the gradient terms that will be generated
in the ultraviolet (UV) through the RG process. Therefore the action
for the field theory describing the problem of interest is given by
\begin{equation}
\mathcal{S}=\mathcal{S}_{\psi}+\mathcal{S}_{\phi}+\mathcal{S}_{\phi\psi}.\label{eq:Action}
\end{equation}
We now proceed with the momentum shell RG accounting for both the
fermionic and bosonic fields.

\section{RENORMALIZATION GROUP}

In this section, we perform one loop RG calculations of the Yukawa
action and derive the flow equations for various coupling constants
in the action. As pointed out in Ref. \cite{Huh,Wang-1}, the presence
of the nodal ring requires that the fermionic and bosonic momenta
be treated differently. While fermionic momenta should be rescaled
towards to the nodal ring, bosonic momenta is rescaled towards the
origin of momentum space. This has important implications for the
tree level scaling analysis.

For fermionic momenta, we take the tree level scaling dimensions to
be 
\begin{equation}
\left[k_{0}\right]=z,\quad\left[\tilde{k}_{r}\right]=1,\quad\left[k_{z}\right]=z_{1},\label{scaling Gpsi}
\end{equation}
where $z_{1}$ is introduced for computational convenience and will
be set to unity later. The scaling dimension of the 3D fermionic integral
$[\int\!dk]=2$ due to the fact that the radius of the nodal ring
$k_{F}$ does not run \cite{Huh,Wang-1}. Tree level scaling invariance
of the fermionic part of the action hence requires that $\left[\psi\right]=-\frac{1}{2}(2z+z_{1}+1)$,
setting the scaling dimensions of velocities as $\left[v_{r}\right]=z-1$
and $\left[v_{r}\right]=z-z_{1}$. For the bosonic part, 
\[
\left[q_{0}\right]=z,\quad\left[q_{x}\right]=\left[q_{y}\right]=\left[q_{z}\right]=1.
\]
Since $[\int\!dq]=3$, this implies that the scaling dimension of
the bosonic fields is $\left[\phi\right]=-\frac{3}{2}\left(z+1\right)$,
whereas the coupling constants $c_{0}$, $c_{1}$ and $c_{2}$ remain
marginal at the tree level. This saves us from unphysical infrared
divergences of the bosonic propagator \cite{Uryszek}. In our analysis,
we assume that important contributions arise when momenta $q$ is
small compared to the radius of the ring $k_{F}$, and hence correspond
to processes with small momentum transfer near the nodal line. In
that spirit, we ignore corrections of the order of $\sim$$|q|/k_{F}$,
as is appropriate when the nodal loop has a large radius compared
to all other energy scales.

To be consistent with this approximation, we have to ensure that one
of the momenta ($q$) in Eq. \ref{eq:Yukawa term} is bosonic, and
the other ($k$) fermionic. With this, one can see that the Yukawa
coupling has scaling dimensions
\begin{equation}
\left[g\right]=\frac{3}{2}\left(z-1\right),\label{g}
\end{equation}
and is therefore marginal at the tree level.

Different implementations of the momentum shell integration have been
employed in the study of nodal-line semimetals. A cylindrical momentum
shell integration scheme was used in Ref. \cite{Huh} whereas a more
symmetric mode elimination in a toroidal geometry was used in Ref.
\cite{Sur-1}. With the understanding that physical quantities do
not depend on the renormalization scheme, we employ a different one
in which frequency and momenta are treated on the same footing. For
fermionic momenta, we perform mode elimination by integrating out
fast modes that lie in a thin shell around the nodal line, 
\begin{equation}
\Lambda e^{-zd\ell}<\sqrt{k_{0}^{2}+E_{k}^{2}}<\Lambda,\label{eq:fint}
\end{equation}
where $E_{k}^{2}=v_{r}^{2}\tilde{k}_{r}^{2}+v_{z}^{2}k_{z}^{2}$.

For bosonic momenta, we integrate out modes that lie in a thin shell
defined by 
\begin{equation}
\Lambda e^{-zd\ell}<\sqrt{q_{0}^{2}+q_{x}^{2}+q_{y}^{2}+q_{z}^{2}}<\Lambda.\label{eq:bint}
\end{equation}
In either case, we assume that the UV energy cutoff $\Lambda$ is
small compared to the radius of the nodal ring, $\Lambda/v_{r}k_{F}\ll1$.
Another important distinction that is specific to nodal-line semimetals
is the fact that in keeping the radius of the nodal line $k_{F}$
fixed, the UV cut-off has a finite scaling dimension $[\Lambda]=-z$,
which is incorporated in the RG flow detailed below.

\begin{figure}[t]
\begin{centering}
\includegraphics[width=0.75\linewidth]{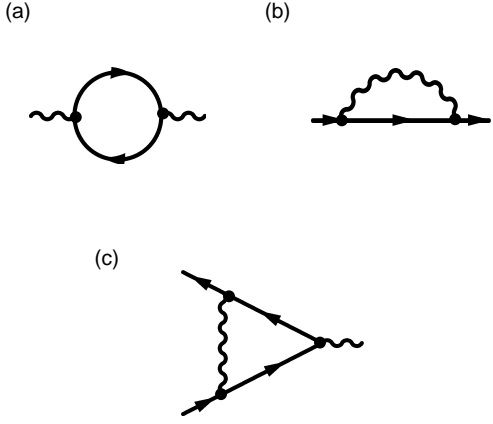}
\par\end{centering}
\caption{{\small{}{}Feynman diagrams at the one loop level. Solid lines represent
the fermionic propagator and dashed lines the bosonic propagator.
(a) Polarization diagram, describing the bosonic self-energy; (b)
fermionic self energy, (c) Vertex corrections.}}
\end{figure}

\subsection{One loop calculations}

At one loop level, corrections to the different coupling constants
in Eq. (\ref{eq:Action}) come from three diagrams: the bosonic polarization
bubble, the fermionic self energy and vertex corrections, as shown
in Fig. 2. The bosonic polarization depicted in Fig. 2a is given by
\begin{equation}
\Pi_{ij}(q)=g^{2}\,\mathrm{tr}\!\int_{>}dk\,\Gamma_{i}G_{\psi}\left(k\right)\Gamma_{j}G_{\psi}\left(k+q\right),
\end{equation}
where $G_{\psi}^{-1}\left(k\right)=-ik_{0}+\mathcal{H}_{0}(k)$ is
the bare fermion propagator
\begin{equation}
G_{\psi}\left(k\right)=\frac{-ik_{0}+v_{r}\tilde{k}_{r}\tau_{0}\otimes\sigma_{y}+v_{z}k_{z}\tau_{0}\otimes\sigma_{x}}{k_{0}^{2}+E_{k}^{2}},\label{Gpsi}
\end{equation}
 $``\text{tr}"$ is the trace and $\int_{>}$ refers to the shell
integration defined in Eq. (\ref{eq:fint}). In all cases, $\Pi_{ij}(q)=\delta_{ij}\Pi(q)$
is diagonal and has the same form for any of the Yukawa vertices considered
in Eq. (\ref{eq:Gammaj}).\textcolor{blue}{{} }To proceed with the integration,
it is convenient to parametrize the fermionic momenta as 
\begin{align}
k_{0} & =\epsilon\cos\theta\nonumber \\
v_{r}k_{r} & =\epsilon\sin\theta\cos\phi\label{eq:parametrization}\\
v_{z}k_{z} & =\epsilon\sin\theta\sin\phi\nonumber 
\end{align}
with $\theta\in\left[0,\pi\right]$ and $\phi\in\left[0,2\pi\right]$
and $\epsilon\equiv\sqrt{k_{0}^{2}+E_{k}^{2}}$ defined within the
UV shell $\epsilon\in[e^{-zd\ell}\Lambda,\Lambda]$ around the nodal
line. As previously stated in the last section, we ignore terms proportional
to $q/k_{F}$ in the denominator of the fermionic propagator, namely
$(|\mathbf{k}_{r}+\mathbf{q}_{r}|^{2}-k_{F}^{2})/2m\approx v_{r}(\tilde{k}_{r}+q_{r}\cos\theta)$,
with $\theta$ the angle between the vectors $\mathbf{k}_{r}$ and
$\mathbf{q}_{r}$. Expanding to second order in $q_{0}$, $q_{r}$
and $q_{z}$ and integrating over $k$, we find that
\begin{equation}
\Pi(q)=\Pi_{0}\left(q_{0}^{2}+\frac{1}{2}v_{r}^{2}q_{r}^{2}+v_{z}^{2}q_{z}^{2}\right)zd\ell,\label{Pi(q)2}
\end{equation}
 where 
\begin{equation}
\Pi_{0}=\frac{4N_{f}g^{2}k_{F}}{3\pi  v_{r} v_{z}\Lambda(\ell)}.\label{Pi0}
\end{equation}
Hence, under the RG process the bosonic action (\ref{Sphi}) is renormalized
as 
\begin{align}
d(c_{0}^{2}) & =\Pi_{0}zd\ell=d(c_{2}^{2})\\
d(c_{1}^{2}) & =\frac{1}{2}\Pi_{0}zd\ell,
\end{align}
whereas the mass term $m_{\phi}$ is not renormalized and can be set
to zero at the fixed point, which corresponds to a quantum phase transition.

We also note that the polarization bubble in the CDW, SC and SDW channels
has explicit frequency dependence, and is distinct from earlier works
that studied the effect of Coulomb interactions in the Yukawa language
by decomposing the four fermion interaction in the Hartree channel
\cite{Huh,Sur-1}. In the Coulomb case, the bosonic propagator is
frequency independent, implying in the absence of renormalization
of the fermionic wavefunction. This in turn implies that vertex corrections
vanish due to Ward identities that follow from the local gauge invariance
of the Yukawa theory\cite{Cho}. The present Yukawa theory is not
locally gauge invariant, and therefore has an intrinsically different
RG structure, with distinct fixed points.

The fermionic self energy in Fig. 2b gives corrections to the velocities
$v_{r}$ and $v_{z}$:
\begin{equation}
\hat{\Sigma}(k)=-g^{2}\sum_{j=1}^{N_{b}}\int_{>}dq\ \Gamma_{i}G_{\psi}(k-q)\Gamma_{i}G_{\phi}(q),
\end{equation}
where
\begin{equation}
G_{\phi}^{-1}\left(q\right)=c_{0}^{2}q_{0}^{2}+c_{1}^{2}(q_{x}^{2}+q_{y}^{2})+c_{2}^{2}q_{z}^{2}\label{Gphi}
\end{equation}
is the bosonic propagator in the disordered phase, and here $\int_{>}$
refers to the shell integration as defined in Eq (\ref{eq:bint}).
Expanding to linear order in $k_{0}$, $\tilde{k}_{r}$ and $k_{z}$,
\[
\hat{\Sigma}(k)=\tau_{0}\otimes\left(-\Sigma_{0}ik_{0}\sigma_{0}+\Sigma_{r}v_{r}\tilde{k}_{r}\sigma_{y}+\Sigma_{z}v_{z}k_{z}\sigma_{x}\right)zd\ell
\]
where 
\begin{align}
\Sigma_{0} & =\frac{N_{b}}{N_{f}}\frac{\Lambda}{v_{r}k_{F}}F_{1}\left(g_{0}^{2},g_{1}^{2},g_{2}^{2}\right)=\Sigma_{z}\label{eq:sigma0}\\
\Sigma_{r} & =\frac{N_{b}}{N_{f}}\frac{\Lambda}{v_{r}k_{F}}F_{2}\left(g_{0}^{2},g_{1}^{2},g_{2}^{2}\right).
\end{align}
The functions $F_{1}$, and $F_{2}$ are special functions, 
\begin{eqnarray*}
F_{1}(a,b,c) & = & \int_{0}^{\pi}\!d\phi\!\int_{0}^{\pi}\!d\theta\frac{\left(2\pi\right)^{-3}\cos2\theta\csc\phi}{\frac{1}{b}+\frac{1}{c}\cot^{2}\phi+\frac{1}{a}\cot^{2}\theta\csc^{2}\phi}\\
F_{2}(a,b,c) & = & \int_{0}^{\pi}\!d\phi\!\int_{0}^{\pi}\!d\theta\frac{\left(2\pi\right)^{-3}\left(\cos2\theta+\cot^{2}\phi\right)\sin\phi}{\frac{1}{b}+\frac{1}{c}\cot^{2}\phi+\frac{1}{a}\cot^{2}\theta\csc^{2}\phi}
\end{eqnarray*}
with 
\begin{equation}
g_{n}^{2}\equiv\frac{N_{f}g^{2}k_{F}}{v_{r} v_{z}c_{n}^{2}\Lambda(\ell)},\quad\text{(}n=0,1,2)\label{a}
\end{equation}
defining a set of three dimensionless couplings.

Lastly, the diagram in Fig. 2c describes the corrections to the Yukawa
vertex,
\begin{equation}
\Upsilon{}_{j}=g^{3}\sum_{i=1}^{N_{b}}\int_{>}dk\,G_{\phi}(-k)\Gamma_{i}G_{\psi}(k)\Gamma_{j}G_{\psi}(k)\Gamma_{i},
\end{equation}
where $\int_{>}$ describes a fermionic momentum shell integration,
as defined in Eq. (\ref{eq:fint}). The calculation of this diagram
yields
\begin{equation}
\Upsilon_{j}=-\frac{(N_{b}-2)}{N_{f}}g\Gamma_{j}\,F_{3}\left(g_{0}^{2},g_{1}^{2},g_{2}^{2}\right)zd\ell.
\end{equation}
where
\[
F_{3}(a,b,c)=\int_{0}^{2\pi}\negmedspace\negmedspace d\phi\!\int_{0}^{\pi}\!d\theta\frac{\left(2\pi\right)^{-3}\csc\,\theta}{(\frac{1}{b}\cot^{2}\theta+\frac{1}{c}\sin^{2}\phi+\frac{1}{a}\cos^{2}\phi)}.
\]
Due to the symmetry of the integrals around the nodal line that follow
from linearly dispersing quasiparticles, no new couplings are generated
in the RG flow. With those results, one can proceed to write down
the RG equations of the problem.

\subsection{RG equations}

Absorbing the renormalizations in the bosonic propagator (\ref{Gphi})
in the form of an anomalous dimension of the bosonic field $\phi$,
\begin{equation}
\eta_{\phi}=\frac{2g_{0}^{2}}{3\pi}z,\label{eta=00005Cphi}
\end{equation}
the $c_{0}$ coefficient becomes marginal whereas $c_{1}$ and $c_{2}$
renormalize as 
\begin{equation}
\frac{d\,\ln c_{1}^{2}}{d\ell}=\frac{2}{3\pi}g_{0}^{2}z\left(\frac{2c_{0}^{2}-c_{1}^{2}}{c_{1}^{2}}\right),\label{c1RG}
\end{equation}
and 
\begin{equation}
\frac{d\,\ln c_{2}^{2}}{d\ell}=\frac{4}{3\pi}g_{0}^{2}z\left(\frac{c_{0}^{2}-c_{2}^{2}}{c_{2}^{2}}\right).\label{c2Rg}
\end{equation}
Regardless of the flow of the dimensionless coupling $g_{0}^{2}$,
it is clear that $\{c_{0}^{2},c_{1}^{2},c_{2}^{2}\}$ flow towards
the values $\{c_{0}^{2},2c_{0}^{2},c_{0}^{2}\}$ at the fixed point,
where $d\ln c_{1}^{2}/d\ell=d\ln c_{0}^{2}/d\ell=0$\textcolor{blue}{,}
provided that $g_{0}^{2}$ remains finite. From Eq. (\ref{a}), the
ratio between the dimensionless couplings at the fixed point is 
\begin{equation}
\frac{g_{1}^{2}}{g_{0}^{2}}=\frac{1}{2},\quad\frac{g_{2}^{2}}{g_{0}^{2}}=1.\label{ratio}
\end{equation}
Since the nature of the interacting fixed point does not depend on
the starting point of the RG flow, we are then allowed to fix the
ratio between $g_{0}$, $g_{1}$ and $g_{2}$ at their fixed point
values in the RG equations from the start \cite{Note}, 
\begin{equation}
g_{0}^{2}=2g_{1}^{2}=g_{2}^{2}\equiv\tilde{g}^{2}.\label{gtilde}
\end{equation}
With this restriction in place, we define the special functions 
\begin{equation}
F_{m}\left(\tilde{g}^{2},\frac{\tilde{g}^{2}}{2},\tilde{g}^{2}\right):=\alpha_{m}(\tilde{g}^{2}),\label{alpha}
\end{equation}
with $m=1,$2,3. Those functions are linear in the dimensionless coupling
$\tilde{g}^{2}$ defined above.

Similarly, one can absorb the RG corrections to the fermionic propagator
by defining the anomalous dimension for the fermionic field $\psi$,
\begin{equation}
\eta_{\psi}=\frac{N_{b}}{N_{f}}\frac{\Lambda}{v_{r}k_{F}}z\,\alpha_{1}(\tilde{g}^{2}),\label{eta_psi}
\end{equation}
in such a way that $dv_{z}/d\ell=v_{z}(z-z_{1})=0$ is marginal if
we set $z_{1}=z$. The remaining RG equation for the other velocity
is
\begin{equation}
\frac{dv_{r}}{d\ell}=v_{r}(z-1)-v_{r}\eta_{\psi}+zv_{r}\Sigma_{r}(\tilde{g}^{2}).\label{eq:betavr}
\end{equation}
 The velocity $v_{r}$ can be kept fixed in the RG flow by renormalizing
the dynamical exponent,
\begin{equation}
z=1+\eta_{\psi}-\frac{N_{b}}{N_{f}}\frac{\Lambda}{v_{r}k_{F}}\alpha_{2}(\tilde{g}^{2})+O(1/N_{f}^{2}).\label{z}
\end{equation}
Ignoring terms that are proportional to $\Lambda/v_{r}k_{F}\ll1$,
consintently with prior assumptions about the fermionic propagator
in Eq. (\ref{Pi(q)2}), we have that $\eta_{\psi}=0$ while the dynamical
exponent $z=z_{1}=1$ remains similarly unchanged at one loop level.
From the preceding analysis, even though $g^{2}$ is marginal, the
tree level scaling dimension of $\tilde{g}^{2}$ is 
\begin{equation}
[\tilde{g}^{2}]=4z-3=1.\label{gtilde2}
\end{equation}
This leads to the one loop RG equation for the dimensionless Yukawa
coupling,
\begin{equation}
\frac{d\widetilde{g}^{2}}{d\ell}=\widetilde{g}^{2}\left[1-\frac{2(N_{b}-2)}{N_{f}}\alpha_{3}(\widetilde{g}^{2})-\frac{2}{3\pi}\widetilde{g}^{2}\right],\label{eq:rgeqn}
\end{equation}
which determines the nature of the interacting fixed point in the
RG flow.

\subsection{Fixed point and critical exponents}

After a quick inspection, the RG equation (\ref{eq:rgeqn}) flows
toward a stable fixed point, where $d\tilde{g}/d\ell=0$. In the $N_{f}\to\infty$
limit, the fixed point is at
\begin{equation}
\tilde{g}_{\infty}^{2}=\frac{3\pi}{2}.\label{ginfty}
\end{equation}
Proceeding in leading $1/N_{f}$ order, the fixed point is
\begin{equation}
\frac{\widetilde{g}_{*}^{2}}{\tilde{g}_{\infty}^{2}}=1-\frac{2(N_{b}-2)}{N_{f}}\alpha_{3}(\tilde{g}_{\infty}^{2}),\label{FP}
\end{equation}
where $\alpha_{3}(\tilde{g}_{\infty}^{2})\approx0.1487$. To compute
the correlation length exponent $\nu$, we write down the RG equations
for the gapping mass term,
\begin{equation}
\frac{dm_{\phi}^{2}}{d\ell}=(2-\eta_{\phi})m_{\phi}^{2},\label{M}
\end{equation}
which gives as a result
\begin{equation}
\nu=\frac{1}{2-\eta_{\phi}}=1-\frac{2(N_{b}-2)}{N_{f}}\alpha_{3}(\tilde{g}_{\infty}^{2}).\label{nu}
\end{equation}

Once two exponents are known, the others can be obtained from hyperscaling
relations. The quantum version of the hyperscaling relation for the
specific heat exponent\textcolor{blue}{{} }gives 
\begin{equation}
\alpha=2-\nu\left(2+z\right)=-1+\frac{2(N_{b}-2)}{N_{f}}\alpha_{3}(\tilde{g}_{\infty}^{2}),\label{eq:alpha}
\end{equation}
\textcolor{black}{where we have set $d=2$ in the general expression
$\alpha=2-\nu\left(d+z\right)$, since the fermions are scaled in
only 2 spatial directions around the nodal line. Moreover, $d=2$
gives the correct large $N_{f}$ behavior, as we show below from the
mean field analysis.}

\begin{table}
\begin{tabular}{cccccccccccccccccccccccccccccccc}
\vspace{-0.2cm} &  &  &  &  &  &  &  &  &  &  &  &  &  &  &  &  &  &  &  &  &  &  &  &  &  &  &  &  &  &  & \tabularnewline
\hline 
\hline 
 &  &  &  &  &  &  &  & Exponent &  &  &  &  &  &  &  & value &  &  &  &  &  &  &  &  &  &  &  &  &  &  & \tabularnewline
\hline 
 &  &  &  &  &  &  &  & $z$ &  &  &  &  &  &  &  & 1 &  &  &  &  &  &  &  &  &  &  &  &  &  &  & \tabularnewline
 &  &  &  &  &  &  &  & $\eta_{\phi}$ &  &  &  &  &  &  &  & $1-0.2975\frac{(N_{b}-2)}{N_{f}}$ &  &  &  &  &  &  &  &  &  &  &  &  &  &  & \tabularnewline
 &  &  &  &  &  &  &  & $\eta_{\psi}$ &  &  &  &  &  &  &  & 0 &  &  &  &  &  &  &  &  &  &  &  &  &  &  & \tabularnewline
 &  &  &  &  &  &  &  & $\alpha$ &  &  &  &  &  &  &  & $-1+0.2975\frac{(N_{b}-2)}{N_{f}}$ &  &  &  &  &  &  &  &  &  &  &  &  &  &  & \tabularnewline
 &  &  &  &  &  &  &  & $\beta$ &  &  &  &  &  &  &  & $1-0.1487\frac{(N_{b}-2)}{N_{f}}$ &  &  &  &  &  &  &  &  &  &  &  &  &  &  & \tabularnewline
 &  &  &  &  &  &  &  & $\gamma$ &  &  &  &  &  &  &  & $1$ &  &  &  &  &  &  &  &  &  &  &  &  &  &  & \tabularnewline
 &  &  &  &  &  &  &  & $\delta$ &  &  &  &  &  &  &  & $2+0.1487\frac{(N_{b}-2)}{N_{f}}$ &  &  &  &  &  &  &  &  &  &  &  &  &  &  & \tabularnewline
 &  &  &  &  &  &  &  & $\nu$ &  &  &  &  &  &  &  & $1-0.2975\frac{(N_{b}-2)}{N_{f}}$ &  &  &  &  &  &  &  &  &  &  &  &  &  &  & \tabularnewline
 &  &  &  &  &  &  &  & \vspace{-0.3cm} &  &  &  &  &  &  &  &  &  &  &  &  &  &  &  &  &  &  &  &  &  &  & \tabularnewline
\hline 
\hline 
 &  &  &  &  &  &  &  &  &  &  &  &  &  &  &  &  &  &  &  &  &  &  &  &  &  &  &  &  &  &  & \tabularnewline
\end{tabular}\caption{List of critical exponents for nodal-line semimetals including $1/N_{f}$
corrections, with $N_{f}$ the number of fermionic flavors and $N_{b}$
the number of components of the bosonic fields. $N_{b}=1$: Charge
density wave order; $N_{b}=2$: superconductivity; $N_{b}=3$: spin
density wave.}
\end{table}

Fisher's and Widom's equality are the same as in the classical case.
Fisher's equality gives the exponent
\begin{equation}
\gamma=(2-\eta_{\phi})\nu=1+O(1/N_{f}^{2}).\label{gamma}
\end{equation}
Essam-Fisher relation $\alpha+2\beta+\gamma=2$ gives the order parameter
exponent in one loop
\begin{equation}
\beta=1-\frac{(N_{b}-2)}{N_{f}}\alpha_{3}(\tilde{g}_{\infty}^{2}).\label{beta}
\end{equation}
Finally, Widom's equality gives the field exponent
\begin{equation}
\delta=1+\frac{\gamma}{\beta}=2+\frac{(N_{b}-2)}{N_{f}}\alpha_{3}(\tilde{g}_{\infty}^{2}).\label{delta}
\end{equation}

The set of exponents for the three different gapped instabilities
and their numerical values is listed in Table I. In nodal line semimetals,
where the existence of a Fermi surface leads the anomalous dimension
of the fermions $\eta_{\psi}$ to vanish in one loop, the only source
of renormalization comes from the vertex correction, which is zero
when $N_{b}=2$. It is clear that in the SC case, where the order
parameter is complex ($N_{b}=2$), all one loop corrections vanish,
implying that the mean field results are exact up to $O(N_{f}^{-2})$
terms. In both the CDW ($N_{b}=1$) and SDW ($N_{b}=3$) order, the
one loop corrections are finite and have opposite signs.

As a consistency check, one can explicitly verify that the mean field
exponents are correctly recovered in the $N_{f}\to\infty$ limit.\textcolor{black}{{}
Taking the bosonic fields at their mean-field value $\phi_{0}$ and
integrating out the fermions,} the mean field free energy for a generic
gapped phase at the nodal line is
\begin{equation}
\mathcal{F}_{\text{MF}}(\phi_{0})=\frac{\phi_{0}^{2}}{g}-\int_{\tilde{k}_{r}^{2}+k_{z}^{2}<\Lambda}d\vec{k}\sqrt{\tilde{k}_{r}^{2}+k_{z}^{2}+\phi_{0}^{2}},\label{MF}
\end{equation}
where $\vec{k}$ are the spatial components of the momentum away from
the nodal line, after conveniently absorbing the velocities $v_{r}$
and $v_{z}$ in their definition. We proceed by expanding (\ref{MF})
in powers of the order parameter $\phi_{0}$ and also in terms of
long-wavelength spatial modulations that couple to momenta as gauge
fields. The corresponding Ginzburg-Landau free energy for nodal-line
semimetals has the usual form expected for conventional Dirac fermions
in 2D \cite{Sachdev},
\begin{equation}
\mathcal{F}_{\text{GL}}(\phi_{0})=\tilde{q}_{r}^{2}|\phi_{0}|+q_{z}^{2}|\phi_{0}|+a_{0}\left(\frac{g_{c}-g}{g}\right)\left|\phi_{0}\right|^{2}+b\left|\phi_{0}\right|^{3},\label{GL}
\end{equation}
where $a_{0}$ and $b$ are positive numbers, and $g_{c}$ is the
critical coupling of the mean field theory. Minimization of the free
energy in the order parameter $|\phi_{0}|$ implies that $\left|\phi_{0}\right|\sim\left|g_{c}-g\right|$
giving $\beta=1$ at the mean field level. At the same time, by dimensional
analysis 
\begin{equation}
\xi^{-2}|\phi_{0}|\cong\delta g|\phi_{0}|^{2},\label{xi}
\end{equation}
and hence the correlation length $\xi\sim(\delta g)^{-1}$ diverges
with the mean field exponent $\nu=1$. Using hyperscaling relations,
all other mean field exponents can be recovered and found to be in
agreement with the large $N_{f}$ results in the $N_{f}$$\rightarrow\infty$
limit, indicating that hyperscaling relations are fulfilled.

\section{CONCLUSIONS}

In summary, we performed a Wilson momentum shell RG calculation and
computed the scaling exponents describing the universal quantum critical
behavior for 3D nodal-line semimetals with linear band crossings.
We considered states that lead to fully gapped instabilities in the
charge, spin and $s$-wave superconducting channels, and calculated
their exponents in a unified manner.

\textcolor{black}{A few comments about the RG literature in nodal-line
semimetals is in order. Previous perturbative RG calculations in Ref.
\cite{Huh,Wang-1} examined the problem of Coulomb interactions within
the Yukawa method for a nodal-line. In those works, a non-interacting
fixed point was found in the clean case, with logarithmic corrections
to scaling, indicating that Coulomb interactions are marginally irrelevant,
as in graphene \cite{Kotov}.}

\textcolor{black}{In the strong coupling regime, Ref. \cite{Sur-1}
addressed the problem of broken symmetry states for a nodal-line both
in the superconducting and in the particle-hole channels. That work
considered the effect of short range interactions through an $\epsilon$-expansion
in the fermionic language. In that approach, short range interactions
are irrelevant operators in the perturbative regime, but flow towards
an interacting fixed point when they are sufficiently strong. The
fermionic language is particularly suitable to address the competition
between different channels of instability, but not so convenient to
address the universal quantum critical scaling of the phases. Here,
we used a non-perturbative Gross-Neveau-Yukawa theory involving order
parameter bosonic fields to examine the quantum critical scaling of
various gapped phases in nodal-line semimetals. After deriving the
interacting fixed points of this theory, we extracted the full set
of quantum critical exponents. Those exponents reduce to their mean-field
values in the $N_{f}\to\infty$ limit, suggesting that hyperscaling
is satisfied.}

The RG calculations were performed in one loop in the number of fermionic
flavors $N_{f}$, which were added for analytic control. \textcolor{black}{We
found that in the SC state, where vertex corrections are absent, the
mean-field exponents are exact within one loop, whereas the CDW and
SDW states have finite $1/N_{f}$ corrections with opposite signs.
In all cases, the dynamical exponent $z=1$ in leading order, whereas
the one loop corrections to various exponents follow directly from
the bosonic anomalous dimension $\eta_{\phi}$}\textcolor{blue}{.}
This study complements current efforts in the literature to account
for the effect of electronic correlations in nodal systems and addresses
a timely class of materials with topological nodal lines.

\section{ACKNOWLEDGEMENTS}

The authors thank R. Nandkishore and F. Kruger for illuminating conversations.
BU acknowledges Carl T. Bush fellowship at University of Oklahoma
for partial support.


\begin{thebibliography}{10}
\bibitem{Note}\textcolor{black}{One could equivalently solve the
RG equations for three independent couplings. That procedure leads
to the same conclusions regarding the fixed point.}
\end{thebibliography}
\end{document}